# CONSTRUCTING AN INVESTMENT FUND THROUGH STOCK CLUSTERING AND INTEGER PROGRAMMING

Maysam Khodayari Gharanchaei, Carnegie Mellon University, Pittsburgh, Pennsylvania, US
Prabhu Prasad Panda, Carnegie Mellon University, Pittsburgh, Pennsylvania, US

**ABSTRACT**

*This paper focuses on the application of quantitative portfolio management by using integer programming and clustering techniques. Investors seek to gain the highest profits and lowest risk in capital markets. A data-oriented analysis of US stock universe is used to provide portfolio managers a device to track different Exchange Traded Funds. As an example, reconstructing of NASDAQ 100 index fund is presented.*

## 1. INTRODUCTION

We aim to create an index fund from a universe of n risky stocks with covariance of returns V. We construct a portfolio that includes at most $k \ll n$ stocks that tracks a specific benchmark portfolio as closely as possible. There are various ways for one to construct a tracking portfolio using significantly less stocks (Anderson et al., 2012) (Ang et al., 2006) (Chow et al., 2011) (Elton and Gruber, 1973). In this project, we used a heuristic approach (Baker et al., 2011) (Frazzini and Pedersen, 2014) (Strongin et al., 2000) as well as an optimization approach (De Carvalho et al., 2012) (Choueifaty, 2008) to mimic three benchmark portfolios (Clarke et al., 2006) composed of $n = 81$ stocks in the NASDAQ composite. From there, we designed portfolios that tracked these three benchmarks using two approaches (DeMiguel et al., 2009) (Frazzini and Pedersen, 2014) (Li et al., 2012). For both heuristic and optimization approaches we use three different values of $k$ stocks with three different rebalancing frequencies (Elton and Gruber, 1973) (Linzmeier, 2011). We measure our turnover (Qian, 2005) and tracking error for each approach and evaluate our results. We discover that with a large value of $k$, we tend to get smaller tracking errors as well as turnover for all the benchmark values we tried out.

We obtained the tickers for NASDAQ 100 through the "pytickersymbols" package and then obtained daily closing prices of the tickers by using the Yahoo Finance package. The data collected ranges from 2014 to 2023. There were a total of 81 stocks, most of which are US-based companies. For each stock, we computed the daily returns to calibrate our model. In order to calculate the market capitalization, we downloaded the quarterly number of shares outstanding from "WRDS Compustat". We split the data to 65% training and 35% testing. The training data set was used to calibrate the model. We chose $k$ stocks out of n stocks based on the Stock Clustering Binary Linear Programming, which used the first 65% of our data starting from 2014.

## 2. OPTIMIZATION APPROACH

Our strategy to form an index fund depends on choosing limited stocks (Strongin et al., 2000) which can represent the index. The most common and practical approach is to choose $k$ stocks, where $k$ is substantially smaller than $n$. The clustering approach was used for picking the $k$ stocks. This approach by default does not guarantee efficient results in terms of mean and variance; it is mainly used to create a portfolio that can track an index as closely as possible. In addition, it forms a partition from our stock universe where stocks are divided up into $k$ clusters, and one stock from each cluster 1 is chosen to represent the other stocks in the cluster. After choosing $k$ stocks to track the index, we followed two strategies to obtain portfolio holdings.

The first strategy is a heuristic approach (Baker et al., 2011) where we found the portfolio holdings based on market capitalization. In this approach, the portfolio holdings are proportional to total market capitalization in each cluster and are normalized so that we can have a fully invested portfolio. The second approach is based on quadratic optimization where we tried to minimize the variance of the portfolio deviation (Chow et al., 2011) to benchmark. We applied usual constraints (Choueifaty, 2008) (Linzmeier, 2011) such as fully invested long only portfolio. In addition, we add boundary for the turnover, as well an equation to make sure we do not assign weights to the stocks that we have not chosen for tracking.

## 3. STOCK CLUSTERING

For the first step, the selection of $k$ stocks to be included in the portfolio can be formulated as a binary linear programming formulation for clustering. We chose correlation between stocks to be a similarity metric in the model. Let the variables be as follows:
- $\rho_{ij}$ is the similarity between stock $i$ and stock $j$,
- $x_{ij}$ indicates which stock $j$ in the portfolio is most similar to $i$. $x_{ij}$ = 1 if $j$ is the most similar stock to $i$ in the portfolio, 0 otherwise,
- $y_i$ indicates which stocks $j$ are in the portfolio. $y_j$ = 1 if $j$ is in the portfolio, 0 otherwise.

Then the formulation is:

$$\max_{\mathbf{x},\mathbf{y}} \sum_{i=1}^{n} \sum_{j=1}^{n} \rho_{ij} x_{ij}$$

$$\text{s.t.} \quad \sum_{j=1}^{n} y_j = k,$$

$$\sum_{j=1}^{n} x_{ij} = 1 \quad \text{for } i = 1, \ldots, n,$$

$$x_{ij} \leq y_j \quad \text{for } i, j = 1, \ldots, n,$$

$$x_{ij}, y_j \in \{0, 1\} \quad \text{for } i, j = 1, \ldots, n.$$

We constructed the Lagrangian formulation

$$L(\mathbf{u}) = \max_{\mathbf{x},\mathbf{y}} \sum_{i=1}^{n} \sum_{j=1}^{n} \rho_{ij} x_{ij} + \sum_{i=1}^{n} u_i \left( 1 - \sum_{j=1}^{n} x_{ij} \right)$$

$$\text{s.t.} \quad \sum_{j=1}^{n} y_j = k,$$

$$x_{ij} \leq y_j \quad \text{for } i, j = 1, \ldots, n,$$

$$x_{ij}, y_j \in \{0, 1\} \quad \text{for } i, j = 1, \ldots, n.$$

By solving this Lagrangian formulation, we can choose a set of $k$ stocks so that each one of the $n$ stocks is similar to one of the $k$ chosen stocks.

## 4. BENCHMARK TRACKING METHODS

We have experimented with three different benchmark indices: equally weighted index, market-cap weighted index, and inverse-volatility constraint.

Our first approach was the equally weighted index. The equally-weighted index would assign $\frac{1}{n}$ weight to each stock within the universe. This would disregard the actual difference between stocks in terms of their size and effect on the market but provides a simple approach to begin with. Mathematically, if $x_i$ denotes stock $i$, then we assign it

$$x_i = \frac{1}{n}, \quad i = 1, 2, \cdots, n$$

Our second method is allocating weight to the stocks corresponding to the ratio of each stock against the entire market capitalization among. Since we have selected our stocks from NASDAQ-100, which actually is calculated using the market capitalization, we believe this approach is more realistic. Since the rebalancing period could occur at any time, we had to match the quarterly data to fit into daily stock return. In Python, we used Pandas library's methods to join the quarterly shares information to the daily return based on the \nearest" optionality. This may at times join the shares outstanding from the future to the past stock return but since the goal of this project is focused on the optimization, we have proceeded to do so. Therefore, at each time we re-balance the portfolio, we were able to calculate the updated benchmark weight at date. Mathematically, if and $x_i$ denotes stock $i$ and $V_i$ denotes the market capitalization of stock $i$, then we assign

$$x_i = \frac{V_i}{\sum_{j=1}^{n} V_j}, \quad i = 1, \ldots, n$$

Our third method was using the inverse-volatility constraint. Inverse volatility is also a value-weighted approach using the inverse volatility as a proxy for the size. When we calculated the volatility, instead of calculating based on the cumulative data, we used a moving-window based approach where the window length is determined by the rebalancing frequency. On top of this, since the weights are determined based on the past window of stock volatility, we had a burn-in period which was solely used to calculate the volatility and was not used to calculate the tracking error. Mathematically, if $x_i$ denotes stock $i$, then we assign it

$$x_i = \frac{1/\sigma_i}{\sum_{j=1}^{n} 1/\sigma_j}, \quad i = 1, \ldots, n$$

where $\sigma_i$ denotes the standard deviation (volatility) of the daily return of asset $i$.

### 5. MARKET CAPITALIZATION APPROACH FOR HEURISTICS

Once the benchmark weight is determined and the stocks were selected from the clustering, we then need to calculate the weight which will be assigned to the selected stocks. Only the selected stocks will have non-zero weights. The weights for the selected stocks will be proportionate to total market capitalization for the cluster represented by that stock normalized the market capitalization of the 81 stocks. Unlike the benchmark weight, for the case of heuristics based approach, we will only be using this single case to assign weight for the selected stocks.

Once the set of k stocks has been selected, we consider the second step of portfolio construction as follows. Assume $j_1, j_2, \ldots, j_k$ are the selected stocks and $C_1, C_2, \ldots, C_k$ are the selected clusters. That is $C_l$ is the set of stocks represented by stock $j_l$ for $l = 1, 2, \cdots, k$. Set the weight of each selected stock $j_l$ proportional to the total market capitalization of the stocks in $C_l$:

$$x_{j_\ell} = \frac{\sum_{i \in C_\ell} V_i}{\sum_{i=1}^{n} V_i}, \quad \ell = 1, \ldots, k$$

where $V_i$ is the market capitalization of stock $i$.
We define the following optimization model to minimize variance and between our portfolio returns and our benchmark returns using only the stocks chosen from the clustering:

$$\begin{aligned}
\min \quad & (\mathbf{x} - \mathbf{x}_B)^T \mathbf{V} (\mathbf{x} - \mathbf{x}_B) \\
\text{s.t.} \quad & \mathbf{1}^T \mathbf{x} = 1, \\
& \mathbf{x} \geq \mathbf{0}, \\
& \|x_t - x_{t-1}\|_1 \leq \text{Turnover bound} \quad \text{for } t = 1, \ldots, n, \\
& x_i \in \{0, 1\} \quad \text{for } i = 1, \ldots, n, \\
& x_i \neq 0,
\end{aligned}$$

where $x_i$ is stock i and $x_B$ is the benchmark portfolio of stocks chosen.
The turnover constraint is:

$$\sum_{i=1}^{n} |x_i^0 - x_i| \leqslant t$$

where $x_0$ and $x$ are respectively a current and new portfolio and $t$ is some threshold.
Our performance metrics are based on the tracking error and the turnover. The tracking error calculated differently for the heuristics and optimization-based approach. The tracking error for the heuristics is calculated by measuring the standard deviation of the difference in returns between the benchmark and our current holdings. The optimization based approach uses the variance calculated as a moving window with length equal to the rebalancing period and weighs it by the difference between the benchmark coefficient and the portfolio coefficient. We considered three different rebalancing frequencies such as quarterly, semi-annually, and yearly. We considered three different values as parameters of the stock clustering model (Linzmeier, 2011) (Frazzini and Pedersen, 2014) (Ang et al., 2006). We repeat these steps for each benchmark (Market Cap Benchmark, Equally Weighted Benchmark, and Inverse Volatility Benchmark).

## 6. RESULTS

### 6.1. Heuristic Approach

After running the heuristic approach on all three benchmarks using $k = 5, 10, 20$ stocks and using quarterly, semi-annually and annually rebalancing periods, we can analyze our results. We noticed a general trend that using more stocks in a tracking portfolio yields lower tracking error as well as turnover for all three benchmarks. Below are graphs of our results from the tracking error of the Market Cap Weighted Benchmark while rebalancing our portfolio quarterly.

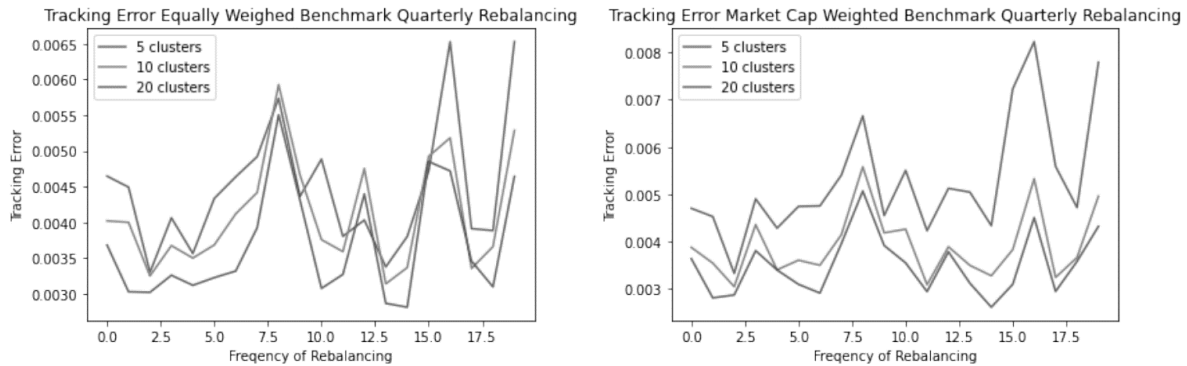

We can also determine that rebalancing our portfolio more frequently generally leads to lower tracking error as well as turnover measure. The tracking error is lower at higher rebalancing frequencies because we have more opportunities to adjust our portfolio to market conditions. The turnover measure is lower because we can re-balance more often and have less positions to rebalances.

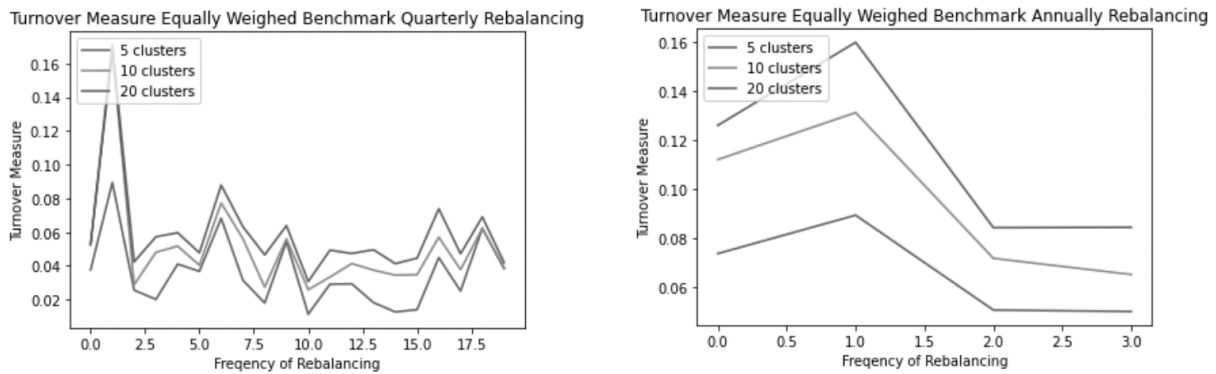

the heuristic approach. We can generally see that the optimization approach yields a much lower tracking error since we have access to much more advanced tools than the heuristic approach.

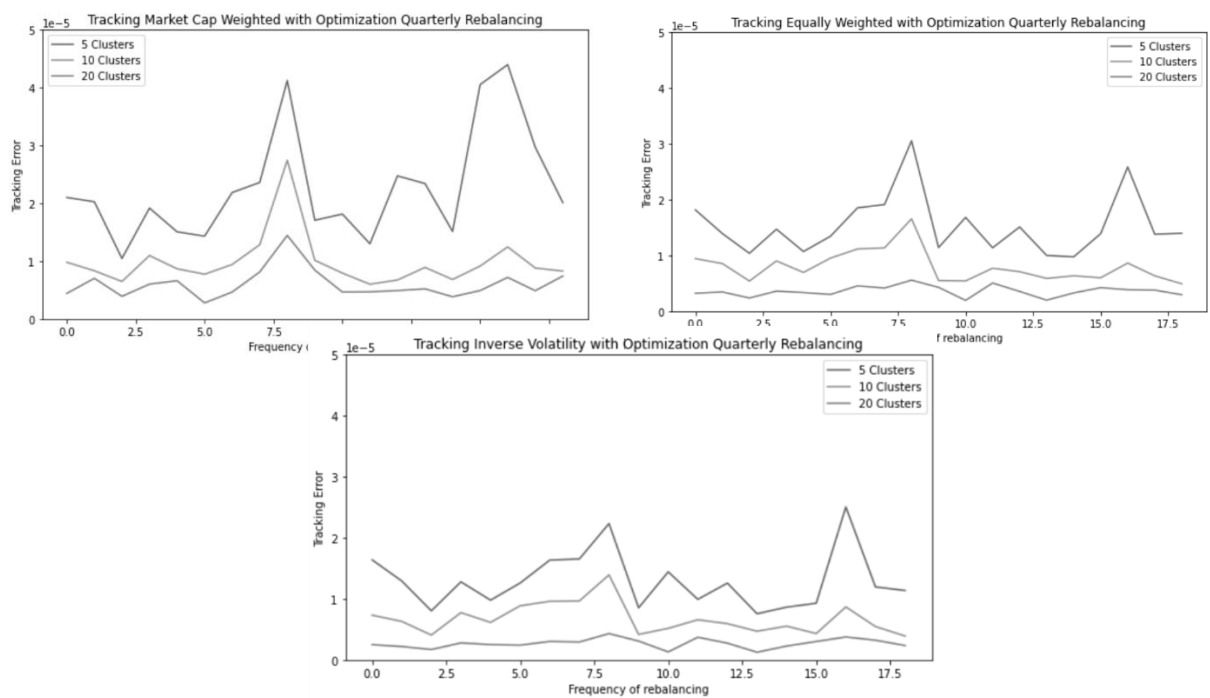

## 7. CONCLUSION

As a conclusion, we followed sequential steps to construct the fund. First of all, we used integer programming to formulate the stock clustering. This is a crucial step in constructing an index fund. We chose correlation as a metric to cluster stocks and we chose stocks based on different values of $k = 5, 10, 20$ to construct our portfolio. The next step was to construct the benchmark. The most simplistic way of constructing a benchmark is to consider an equally weighted portfolio, we also constructed a benchmark based on inverse variance and market capitalization. After constructing a benchmark, we followed two approaches to construct our index. The first approach assigned weights to our portfolio based on market capitalization, and the second approach was to optimize weights based quadratic programming with added constraints related to turnover. We tried three rebalancing frequencies such as quarterly, semi annually, and annually. As such, we discovered that with a large value of k, we tend to get smaller tracking errors as well as turnover for all the benchmark values we tried out. Furthermore, the optimization approach yields a much lower tracking error since we have access to much more advanced tools than the heuristic approach.

## 8. References and Bibliography

ANDERSON, R. M., BIANCHI, S. W. & GOLDBERG, L. R. 2012. Will my risk parity strategy outperform? *Financial Analysts Journal,* 68**,** 75-93.
ANG, A., HODRICK, R. J., XING, Y. & ZHANG, X. 2006. The cross-section of volatility and expected returns. *The journal of finance,* 61**,** 259-299.
BAKER, M., BRADLEY, B. & WURGLER, J. 2011. Benchmarks as limits to arbitrage: Understanding the low-volatility anomaly. *Financial Analysts Journal,* 67**,** 40-54.
CHOUEIFATY, Y. 2008. Towards maximum diversification. *Available at SSRN 4063676*.
CHOW, T.-M., HSU, J., KALESNIK, V. & LITTLE, B. 2011. A survey of alternative equity index strategies. *Financial Analysts Journal,* 67**,** 37-57.
CLARKE, R., DE SILVA, H. & THORLEY, S. 2006. Minimum-variance portfolios in the US equity market. *Journal of Portfolio Management,* 33**,** 10.

DE CARVALHO, R. L., LU, X. & MOULIN, P. 2012. Demystifying equity risk–based strategies: A simple alpha plus beta description. *The Journal of Portfolio Management,* 38**,** 56-70.
DEMIGUEL, V., GARLAPPI, L. & UPPAL, R. 2009. Optimal versus naive diversification: How inefficient is the 1/N portfolio strategy? *The review of Financial studies,* 22**,** 1915-1953.
ELTON, E. J. & GRUBER, M. J. 1973. Estimating the dependence structure of share prices--implications for portfolio selection. *The Journal of Finance,* 28**,** 1203-1232.
FRAZZINI, A. & PEDERSEN, L. H. 2014. Betting against beta. *Journal of financial economics,* 111**,** 1-25.
LI, X., SULLIVAN, R. & GARCÍA-FEIJÓO, L. 2012. The limits to arbitrage revisited: the low-risk anomaly. *Financial Analysts Journal*.
LINZMEIER, D. 2011. Risk balanced portfolio construction. Working paper.
QIAN, E. E. 2005. On the financial interpretation of risk contribution: Risk budgets do add up. *Available at SSRN 684221*.
STRONGIN, S., PETSCH, M. & SHARENOW, G. 2000. Beating benchmarks. *The Journal of Portfolio Management,* 26**,** 11-27.